# Completing the puzzle: why studies in non-human primates are needed to better understand the effects of non-invasive brain stimulation

**Authors and Affiliations:**

Sebastian J. Lehmann (1,*), Brian D. Corneil (1,2,3)

Department of Physiology and Pharmacology, Western University, London, Ontario, Canada, N6A 5B7

Department of Psychology, Western University, London, Ontario, Canada, N6A 5B7

Robarts Research Institute, London, Ontario, Canada, N6A5B7

**(*) Corresponding Author:** Sebastian J. Lehmann ( slehmann@uwo.ca )

Co-Author: Brian D. Corneil ( bcorneil@uwo.ca )

**Graphical abstract:** Brain stimulation - facets of a multidisciplinary field

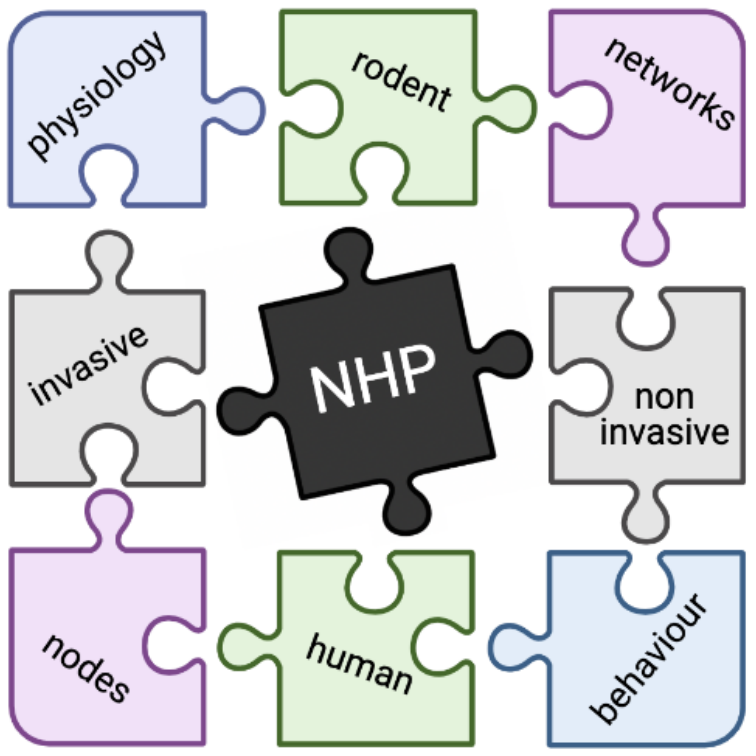

**Acknowledgments:**

Figures were created with "biorender.com"

## Highlights

- Despite widespread use in the lab and clinic, it remains unclear how non-invasive brain stimulation (NIBS) modulates neural activity in targeted areas and networks, and how such modulation ultimately impacts behaviour
- Non-human primates (NHPs) offer a unique animal model to help bridge this gap in knowledge
- Due to its homology and accessibility, the NHP oculomotor system that moves the line of sight offers a particularly valuable system for revealing neurophysiological insights during and after NIBS that generalize to humans
- Establishing the NHP as an animal model for NIBS promises a better understanding of the underlying mechanisms, which may ultimately advance the treatment of complex cognitive disorders

## Abstract

Brain stimulation is a core method in neuroscience. Numerous non-invasive brain stimulation (NIBS) techniques are currently in use in basic and clinical research, and recent advances promise the ability to non-invasively access deep brain structures. While encouraging, there is a surprising gap in our understanding of precisely how NIBS perturbs neural activity throughout an interconnected network, and how such perturbed neural activity ultimately links to behaviour. In this review, we will consider why non-human primate (NHP) models of NIBS are ideally situated to address this gap in knowledge, and will consider why the oculomotor network that moves our line of sight offers a particularly valuable platform in which to empirically test hypothesis regarding NIBS-induced changes in brain and behaviour. NHP models of NIBS will enable investigation of the complex, dynamic effects of brain stimulation across multiple hierarchically interconnected brain areas, networks, and effectors. By establishing such links between brain and behavioural output, work in NHPs can help optimize experimental and therapeutic approaches, improve NIBS efficacy, and reduce side-effects of NIBS.

## 1. Introduction

The past decades have seen remarkable advances in our ability to perturb brain activity in humans. Whether it is deep brain stimulation (DBS) delivered invasively to the basal ganglia of patients with Parkinson's disease (PD) or transcranial magnetic stimulation (TMS) delivered non-invasively to the motor cortex of a research subject, brain stimulation continues to be an essential tool in human neuroscience. Currently, there are numerous non-invasive brain stimulation (NIBS) techniques that are established or emerging (Polanía et al., 2018), such as TMS, transcranial electric stimulation (tES, including direct [tDCS], alternating current [tACS], and temporal interference stimulation [TI]), or focused ultrasound stimulation (fUS). Stimulating the human brain offers the opportunity to make causal inferences about brain function, or modulate neural activity for clinical therapy, with the potential to alleviate disease burden in numerous conditions where patients become unresponsive to first- or second-line therapeutics (Blumberger et al., 2018; Grover et al., 2021). Other emerging non-invasive approaches include neuromodulatory techniques such as vagal nerve stimulation (Bhattacharya et al., 2021), or fUS to enable temporary and accurate opening of the blood-brain barrier for targeted drug delivery to the central nervous system (Abrahao et al., 2019).

Despite the widespread use of NIBS in the lab and clinic, there is a substantial gap in our understanding of exactly how altered neural activity consequent to NIBS ultimately impacts behaviour. For established techniques like TMS, this gap arises not necessarily from a lack of knowledge about the immediate underlying biophysics that modulate neural activity, but rather from uncertainties in how such induced neural activity influences activity within an interconnected network, and how such network activity influences behaviour. These uncertainties do not come as a surprise, since they are present even on a small network scale when using optogenetics, which can precisely modulate activity of specific cell types and circuits (Miesenböck, 2011). Such uncertainties also accompany other forms of NIBS (e.g., tES, fUS) in which key questions about the induced local and global network effects remain to be established (Polanía et al., 2018). Further, both basic and clinical applications of NIBS are plagued with substantial inter-subject variability (Hamada et al., 2013; Hordacre et al., 2017; Huang et al., 2017; Lefaucheur et al., 2014; Ridding & Ziemann, 2010). While a comprehensive mechanistic understanding of how NIBS works is not a prerequisite for therapeutic applications, such an understanding promises improved efficiency, a reduction in side-effects, and potentiates the development of new NIBS implementations in both clinical practice and research. While the safety of some forms of NIBS like TMS (Boes et al., 2018) or tES (Antal et al., 2017) are well established, knowledge of the safety of more recently developed techniques like fUS is rather limited, and further investigations in multiple species are needed to establish a solid safety framework in order to prevent side effects and neuronal damage (Pasquinelli et al., 2019).

The purpose of this review is to illustrate how work in animal models, and specifically in non-human primates (NHPs), can play a crucial role in addressing how modulated neural activity following NIBS ultimately impacts behaviour. We will place a particular emphasis on TMS as an exemplar NIBS technique, as this technique enjoys widespread use. We will also focus on a series of neurophysiological findings in the oculomotor network that moves our lines of sight. The oculomotor network is arguably the best-understood sensorimotor network in the primate brain, hence this network offers a platform in which to test theories of NIBS function, and to compare such NIBS results to those arising from more direct manipulations of neural activity (e.g., intracortical microstimulation, or pharmacological manipulation). Such comparisons are essential given that ideas about NIBS often propose simplified heuristics (e.g., a “virtual lesion”, or “rebalancing activities across cortical hemispheres”) that do not detail the mechanism of action, capture the dynamics of how induced changes in neural activity reverberate throughout an interconnected network, nor link such changes to behaviour when NIBS is applied outside of primary sensory or motor areas. Although this review will be primarily on TMS and its effects on the oculomotor network, we hope the concepts being raised, and the way in which NHP models can help fill these gaps in knowledge, will generalize to other forms of NIBS.

## 2. Probing the brain using NIBS

### 2.1. A brief history of TMS

TMS was first introduced in the 1980s (Barker et al., 1985), and it has evolved to one of the most frequently used NIBS techniques in basic and clinical research. The biophysics and contemporary use of TMS have been covered in detail elsewhere (e.g., Hallett, 2007). Very briefly, in TMS a rapidly changing magnetic field induces an electric field in the brain which, if of sufficient amplitude and duration, initiates action potentials at excitable target areas (Hallett, 2007; Rossini et al., 2015). Biophysical models suggest that single pulses of TMS perturb ca. 1-2 square centimeter of cortical tissue in the human brain, although this depends on many factors such as stimulation intensity, subject-specific brain anatomy, coil-orientation, and coil design (Bergmann et al., 2016; Deng et al., 2014; Weise et al., 2020). Single-pulses of TMS can provoke percepts when applied over primary visual areas (Fried et al., 2011; Schaeffner & Welchman, 2017), or induce motor evoked potentials (MEPs, or brief twitches of muscle activity) when applied over primary motor cortex (M1) of humans and macaques (Baker et al., 1994; Barker et al., 1985). MEPs in particular provide an objective measure of the effects of TMS, and often serve as the dependent measure of interest in many studies.

A variety of TMS protocols enjoy widespread use, and can be broadly subdivided into online and offline approaches (Bergmann & Hartwigsen, 2020). Online TMS protocols involve the precisely-timed delivery of one or more TMS pulses during the performance of a behaviour task. By observing the effects of such intermittent pulses on performance, one can infer the contribution of the stimulated area to the behavior in question. Online TMS protocols often employ the logic of state-dependency, where the effects of TMS are augmented if the stimulated area is engaged at the time of TMS, presumably due to interactions between TMS pulses and endogenous activity present at the time of stimulation. State-dependent strategies can also include the delivery of stimulation pulses timed to endogenous brain frequencies (Bergmann et al., 2019; Stefanou et al., 2019; Thut et al., 2017). In addition to targeting single nodes, multifocal approaches such as cortico-cortical paired associative stimulation protocols repeatedly deliver pairs of pulses to interconnected areas in order to selectively strengthen or disrupt functional connectivity (Buch et al., 2011; Fiori et al., 2018; Johnen et al., 2015).

During a typical offline TMS protocol, one or a sequence of TMS pulses are applied repeatedly while a subject is quiescent. A variety of modes of such repetitive TMS (rTMS) are available, e.g., 1- or 5-Hz rTMS, continuous (cTBS) or intermittent theta-burst stimulation (iTBS); for an overview see Klomjai et al., 2015; Veniero et al., 2019). The presumed physiological effect of such rTMS protocols is largely based on its influence on the MEP when rTMS is applied over M1. Protocols that decrease MEP magnitude are often said to produce a “virtual lesion” of the stimulated area, whereas those that increase MEP magnitude are thought to facilitate the stimulated area’s output (Huang et al., 2005; Pascual-Leone et al., 2000). Such observations often guide the use of TMS for therapeutic effect, since many neurodevelopmental disorders have been linked to alteration in the balance of excitatory and inhibitory influences within key brain networks (Klomjai et al., 2015). For example, repetitive modes of TMS are often used to try to re-balance activity across the hemispheres in the context of stroke or depression, by either reinforcing an underactive or damaged area through facilitatory rTMS protocols, or by reducing the output of an overactive mirroring homotopic area through inhibitory rTMS protocols (Downar & Daskalakis, 2013; Ridding & Rothwell, 2007; Takeuchi et al., 2005).

## 2.2. Bridging the gap between brain and behaviour

One central goal of cognitive neuroscience is to generate and test mechanistic theories of behaviour. Brain stimulation techniques such as TMS enable causal tests of such theories in a variety of ways, often following a “virtual lesion” logic where the contribution of a given area is inferred from the ability of stimulation to disrupt behavior. This approach was first adopted for studies in M1, with single-pulses of TMS-M1 evoking smaller MEPs when applied after a train of rTMS-M1 pulses (Chen et al., 1997), and was soon extended to percepts evoked from primary sensory areas (Fried et al., 2011). While a convenient shorthand, the implied analogy to actual lesions following focal brain damage in patients or experimental ablation of tissues can be misleading. Subsequent work in M1 showed that different patterns of rTMS could either facilitate or inhibit MEPs (Huang et al., 2005, 2017), hence the changes in evoked responses after rTMS better parallel long term potential or depression phenomena induced by electrical stimulation. Regardless of the best analogy, generalizing results and approaches beyond primary or sensory motor areas is fraught with a number of potential problems, first and foremost the lack of an immediate, quantifiable, and reliable overt response. A recent brain-mapping study in humans confirmed the robustness and simplicity of percepts evoked from primary sensory areas, but showed that induced percepts were both more complex and harder to evoke from higher-order brain areas (Fox et al., 2020). For TMS-M1, the MEP is conveyed via corticospinal pathways, but M1 is unique in regards to both its intrinsic circuitry, connections with other brain areas, and its efferent projections (Polanía et al., 2018; Rossini et al., 2015). Leaving aside whether stimulation thresholds derived from M1 can generalize to other brain areas, TMS intensities capable of provoking quantifiable responses increase when applied outside of primary sensory or motor areas (Davare et al., 2006; Fried et al., 2011), presumably increasing the volume of activated tissues and lowering the precision with which a given effect can be ascribed to a particular area. Furthermore, the effects of a given protocol can be remarkably variable depending on the targeted area. For example, a recent study combining TMS and imaging reported how an ostensibly 'inhibitory' rTMS protocol decreased local inhibition and disrupted feedforward and feedback connections when applied to frontal cortex, but increased local inhibition and enhanced feedforward signalling when applied to occipital cortex (Castrillon et al., 2020).

In addition to the concerns and limitations in trying to generalize protocols or approaches from M1 to other brain areas, TMS-M1 results can themselves be remarkably variable.

For example, Hamada and colleagues examined the MEPs to TMS-M1 following putative inhibitory or excitatory rTMS protocols. While only half of the sample showed the expected changes in MEPs, the authors found that those who exhibited unexpected results tended to also have longer latency MEPs; the authors attributed these differences to the interneuron network recruited by TMS-M1 (Hamada et al., 2013; Hordacre et al., 2017). These results hint at an underlying anatomical and physiological basis to the variability inherent to results from TMS-M1. Unique anatomical and physiological factors will almost certainly impact the results when TMS is applied to other brain areas. Further, brain networks themselves are highly dynamic, changing at short times scales and depending on the immediate history of neural activity (Karabanov et al., 2015; Silvanto et al., 2009; Stefanou et al., 2019). In addition to short term variation of activity on a local or behaviour-specific network level, the variability of effects following NIBS may be accounted to a slow drift of global brain states, which is often - but not exclusively - reflected in covert processes like attentional changes or impulsivity, potentially affecting changes in overt behaviour  (Cowley et al., 2020).

In order to help understand and potentially reduce the variability present in NIBS results, recent work in the human literature has seen extensive efforts combining NIBS with non-invasive forms of neuroimaging like EEG, MEG, fMRI or fNIRS (Bergmann et al., 2016; Thut et al., 2017). These efforts are essential to help better understand the endogenous state of the brain at the time of stimulation, and can also help detail the effects of NIBS on the functional connectivity of a targeted brain network. Other work has used non-invasive imaging to optimize NIBS in a subject-specific and functionally-dependent manner (Bergmann et al., 2016; Weise et al., 2020), which can for example help define the behaviourally-relevant variations in structure and function within the hand motor area (Dubbioso et al., 2021). More broadly, the combination of NIBS and neuroimaging offers the opportunity to bring brain activity in specific areas and networks under transient experimental control, and tailoring this approach to subject-specific endogenous brain activity permits closed-loop TMS triggered by oscillatory markers of excitability (Bergmann et al., 2019; Schaworonkow et al., 2019; Schilberg et al., 2018; Stefanou et al., 2019) or frequency-tuned short- and long-term oscillation entrainment via repetitive modes of TMS (Hanslmayr et al., 2014; Thut et al., 2011, 2012) or tES (Herrmann et al., 2013; Neuling et al., 2013; Zaehle et al., 2010). Doing so can enable causal tests of this aspect of brain function on behaviour on the local and network level (for a comprehensive review see Thut et al., 2017). Therefore, such approaches also incorporate the fields of neuromodulation and brain connectomics (as introduced by Sporns et al., 2005). Rather than relying on the average interareal connectivity on the population level (normative connectomics), individualized knowledge about network connectivity can provide further opportunities to understand and modulate subject-specific network processes (Horn & Fox, 2020). Using such individualized approaches, the structural and functional connectivity of targeted brain areas can serve as predictors of therapeutic success for DBS of the subthalamic nucleus in Parkinson's Disease (Horn et al., 2017) and rTMS of PFC in depression (Cash et al., 2019), further illustrating the importance of understanding the link between brain networks and behaviour in order to optimize efforts for therapeutic applications (Ridding & Ziemann, 2010; Ziemann & Siebner, 2015).

## 2.3. Remaining gaps in knowledge

Despite this progress in improving the delivery and application of NIBS in humans, a considerable number of questions and gaps remain. Our understanding of how activity within, and communication between, nodes of an interconnected network give rise to behaviours is still lacking; such an understanding remains vital as a basis in which to interpret the perturbations caused by NIBS. This is even more important since null-results for disrupting structure-function relationships are not necessarily a proof for the absence of such a relation. Relatedly, non-invasive imaging techniques in humans have inherent limitations regarding spatial and temporal resolutions; while some of them lack the ability to acquire information from subcortical nodes of interconnected networks, others with access to such nodes are also not entirely informative about the direction of connectivity between brain areas (Horn & Fox, 2020). Therefore, research investigating the physiological and behavioural effects of NIBS will benefit from a comparison of such results with those obtained using more invasive and precise approaches, such as intracortical microstimulation, optogenetic manipulation, temporary lesioning, or permanent ablation. Combining non-invasive stimulation techniques with invasive neural recording techniques can help link the effects of stimulation to behavioral consequences. In the following sections we will argue why the use of animal models, in particular non-human primates (NHPs) trained to perform complex tasks that emulate aspects of human behaviour, can play an essential role in helping to fill in these gaps in knowledge and ultimately form a crucial impact on both basic and clinical research in human populations.

# 3. The role of NHP animal models in NIBS research

## 3.1. Animal models for TMS

While the vast majority of TMS research has been conducted in human subjects and patients, in vivo and in vitro work on animal models has contributed significantly to the understanding of the immediate and long-term neurophysiological effects induced by TMS (Tang et al., 2017). Work in mouse hippocampal cultures (Lenz et al., 2015; Vlachos et al., 2012), acute rat brain slices (Pashut et al., 2014) or intact, behaving rodent models (Makowiecki et al., 2014; Mix et al., 2010; Rodger et al., 2012; Tang et al., 2017) and cats (Allen et al., 2007; Kozyrev et al., 2014, 2018) have led to important mechanistic insights at the synapse and microcircuit levels, detailing long-term depression and potentiation phenomena following rTMS protocols. However, a direct translation of such results to the human brain, and ultimately to behaviour, is uncertain. The brains of mice, rats, and cats have evolved to meet distinct evolutionary niches, and numerous recent reviews have detailed important differences between whole-brain connectivity of the most used mammalian animal models in neuroscience (Garner, 2014; Hutchison & Everling, 2012; Schaeffer et al., 2020). While work in mice and rats offers the extraordinary opportunity to combine NIBS with invasive experimental techniques, the impact of state dependency on the results produced by NIBS means that it is essential to conduct parallel studies in awake, higher-order animal models capable of generating complex, human-like behaviours. Doing so provides further opportunities to better understand how induced activity within interconnected networks gives rise to behaviours.

### 3.2. Monkey in the middle – the role of NHPs as a model of higher cognitive function

Given its phylogenetic proximity to humans, the NHP, and specifically the old-world rhesus macaque, is a valuable and well-established animal model for higher cognitive function in humans. Comparative studies of brain networks in humans and NHPs continue to emphasize a high degree of anatomical and functional homology (Bullmore & Sporns, 2009; Hutchison & Everling, 2012), and the macaque brain exhibits a greater degree of gyrification than other animal models. Such gyrification is important in that it introduces complexities to the underlying biophysics of stimulation that are present in humans. Early studies in NHPs confirmed that TMS-M1 pulses activated corticospinal neurons, reporting that the induced responses varied depending on cell-body-size, axonal conduction velocity and trajectory, as well as by the location of neurons and their excitability compared to stimulation threshold, which varied when identifying direct (D) and indirect (I) responses (S. Edgley, 1997). Edgley and colleagues (S. A. Edgley et al., 1997) also found that both TMS and TES evoked monosynaptic responses in corticospinal neurons, helping to define the biophysics of induced responses to NIBS in the primate brain. Subsequent work confirmed that the macaque is a suitable model in which TMS could be applied during the performance of a variety of complex behavioural tasks involving eye or manual responses (Balan et al., 2017, 2019; Gu & Corneil, 2014; Merken et al., 2021; Mueller et al., 2014; Ortuno et al., 2014; Romero et al., 2019), and that such approaches can be combined with invasive neural recordings in the targeted area (Mueller et al., 2014; Romero et al., 2019). Furthermore, behaving NHP models are almost amenable to other forms of NIBS such as tES (Johnson et al., 2020; Krause et al., 2017, 2019; Opitz et al., 2016; Vieira et al., 2020) and focused ultrasound (Deffieux et al., 2013; Kubanek et al., 2020; Verhagen et al., 2019; Wattiez et al., 2017; Yang et al., 2018, 2021).

# 4. The primate oculomotor system

## 4.1. Anatomy and physiology of the primate oculomotor system

Humans and other primates are largely visual animals; large swaths of the cortex are devoted to processing and understanding the visual image. The foveate nature of primate vision, where high-resolution vision is realized on a restricted portion of the retina, necessitates an oculomotor system capable of either stabilizing an object of interest on retina, or rapidly repositioning the eyes toward a new object of interest. Humans and non-human primates (NHPs) exhibit a very similar oculomotor repertoire, including the generation of saccadic movements that rapidly re-orient the line of sight. Studies of saccades in either humans or NHPs are highly suited to address the question about how NIBS-induced changes influence neural activity and behaviour, given the comparative ease with which visual input can be controlled and eye movements can be measured.

The oculomotor system is distributed across numerous subcortical and fronto-parietal nodes that respectively encompass both the low-level machinery that rapidly moves the eyes, and the higher-level circuits that implement the flexible strategies necessary to operate efficiently in a complex and dynamic environment (Bisley & Goldberg, 2010; Corneil & Munoz, 2014; Krauzlis et al., 2013; Sommer & Wurtz, 2008) (*Figure 1A*). Similarities between macaques and humans extend to high-level cognitive abilities; rhesus macaques appear unique amongst animal models used in neuroscience in their ability, like humans, to remember thousands of visual images after only a single viewing (Meyer & Rust, 2018). The NHP oculomotor system has been intensively studied for over 50 years, yielding a refined understanding of the immediate premotor and motor events preceding a saccade (Bisley & Goldberg, 2010; Corneil & Munoz, 2014; Hanes & Schall, 1996; Krauzlis et al., 2013; Sommer & Wurtz, 2008). Complementary work with non-invasive imaging techniques in humans (e.g., EEG, fMRI, MRI) continues to emphasize the high degree of anatomical and functional homology in the oculomotor system of humans and NHPs (Ford et al., 2009; J. F. Mitchell & Leopold, 2015; Petit & Pouget, 2019; Schaeffer et al., 2020). At the level of the brainstem reticular formation, saccade execution is governed by a balance of mutually-antagonistic populations of fixation-related omni-pause neurons (OPNs) and saccade-related burst neurons (Scudder et al., 2002; Sparks, 2002). While a complete picture of precisely how saccade threshold is implemented within the oculomotor brainstem is still lacking (Jantz et al., 2013; Peel et al., 2017; Schall, 2019), important inputs come from the midbrain superior colliculus (SC) and frontal eye fields (FEF), both of which also contain mutually-antagonistic populations of saccade- and fixation-related neurons (Dorris & Munoz, 1995; Izawa et al., 2009; Schall, 2013). The SC, by virtue of its unique receipt of converging inputs from fronto-parietal cortex and outputs to premotor structures, provides an interface by which myriad variables can be integrated into a saccadic motor plan. Within the SC, FEF, and other frontal and parietal oculomotor areas, one finds neural activity that correlates with any number of sensory, cognitive, or motor variables (Freedman & Ibos, 2018; Sommer & Wurtz, 2008; Tehovnik et al., 2000), although sensory and cognitive processes are more heavily weighted in the cortical nodes, whereas motor-execution signals are more heavily weighted in the SC. The representation of this sensorimotor continuum across nodes reflects the high degree of reciprocal connectivity within the network. Importantly, communication amongst these nodes is not simply “top-down”, as all structures are inter-connected, and subcortical structures such as the superior colliculus can influence cortical processing via signals relayed through the thalamus (Basso et al., 2021; Cavanaugh et al., 2020; Shine, 2020; Sommer & Wurtz, 2008) .

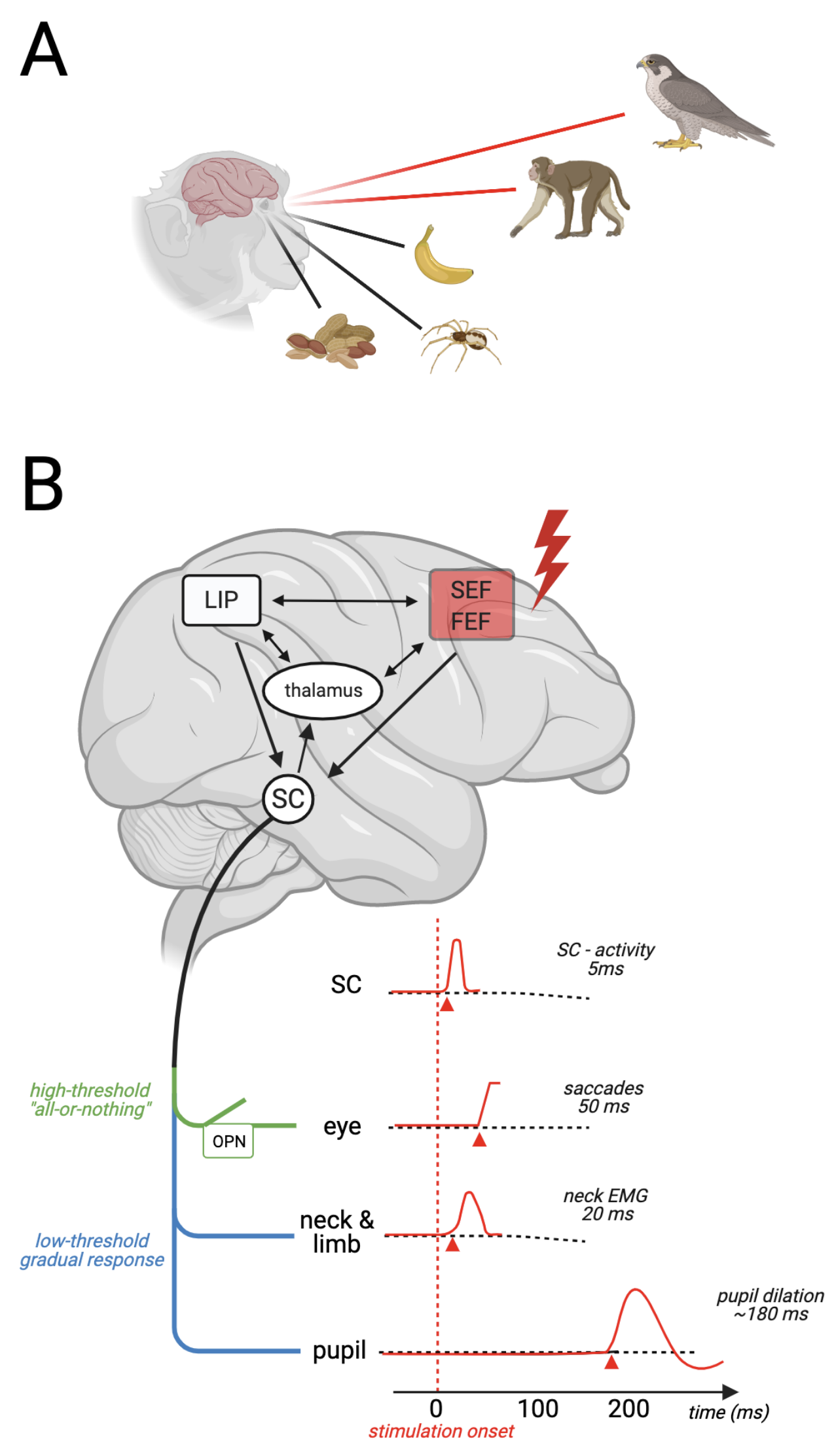

***Figure 1 A. The Orienting Response - interacting with a dynamic environment.*** *Components of the orienting response are essential for survival in a dynamic environment. They include overt responses, shifting the line of sight towards targets, as well as covert attentional processes, allowing for goal-directed interaction during foraging and social interaction, as well as for coordinating fast responses to imminent threats from predators.* ***B. From brain to behaviour: neural substrates and effectors of the primate orienting system****. The orienting response is mediated via the oculomotor system, a network distributed across cortical (e.g., FEF, LIP) and subcortical brain areas (e.g., SC, thalamus). Components of the response can be measured across a variety of downstream effectors, including the eyes (saccades, changes in pupil diameter) as well as neck, shoulder, and limb muscles. Physiological responses to salient visual or auditory stimuli can be measured across those effectors, revealing effector-specific, time-locked responses. Artificially stimulating areas of the oculomotor system like the SC or FEF can trigger such responses (red curves, arrows indicate response onset).*

In NHPs, the oculomotor system is most commonly studied via saccadic eye movements made with the head restrained. However, the motor output of the primate SC is an orienting command that is distributed to many effectors (Corneil & Munoz, 2014; Gandhi & Katnani, 2011) (*Figure 1B*). Anatomical work revealed that SC efferents not only project to brainstem saccade circuits, but also to premotor centers for head, limb, and autonomic control (Grantyn & Grantyn, 1982). There is good evidence that the potent inhibition of brainstem OPNs applies only to the discharge of saccadic premotor burst neurons, and not to the other components of the orienting response. For example, subthreshold levels of electrical stimulation within the SC that are below the level to evoke saccades can nevertheless provoke orienting neck muscles responses (Corneil et al., 2002) or pupil dilation (Wang et al., 2012). SC activity related to the presentation of visual targets, or cognitive variables such as the allocation of visuospatial attention or reward can also elicit neck muscle activity (Corneil et al., 2004, 2008; Rezvani & Corneil, 2008), and sub-saccade threshold stimulation of the FEF was shown to evoke pupil dilation (Lehmann & Corneil, 2016), the recruitment of a head turning synergy (Corneil et al., 2010), or the allocation of visuospatial attention (Ebitz & Moore, 2017; Moore & Fallah, 2004).

### 4.2. Non-invasive perturbation of the primate oculomotor system

Given their superficial location and accessibility, the parietal and frontal oculomotor areas are common targets for non-invasive stimulation experiments in both humans and NHPs. While invasive intracortical stimulation of these areas can elicit saccadic eye movements (Bruce et al., 1985; Mushiake et al., 1999), attempts to evoke such movements with TMS have been largely unsuccessful (Müri et al., 1991; Wessel & Kömpf, 1991). Unlike the motor cortex, neither frontal nor parietal oculomotor areas directly project to motoneurons, hence any effect of NIBS has to be relayed through the brainstem (Stanton et al., 1988a, 1988b). In the absence of a direct output measure, researchers have assessed how TMS of those frontal (Müri et al., 1991; Priori et al., 1993; Thickbroom et al., 1996; Valero-Cabre et al., 2012) and parietal areas (Silvanto et al., 2009) influence a variety of oculomotor or cognitive behaviours (see Vernet et al., 2014 for a review focused on frontal cortex). While it is established that NIBS to frontal and parietal oculomotor areas can influence such behaviors, there is little to no consensus as to the underlying mechanism. In part, this reflects the heteromodal nature of frontal and parietal oculomotor areas: they are higher-level areas whose contributions to complex behavioural tasks, let alone the responses of such areas to stimulation during such tasks, are still being determined. However, mechanistic uncertainty also arises from the dichotomous nature of oculomotor control. Consider for example a situation where TMS increases saccadic reaction time (RT) in a given behavioural task. Such an effect can be attributed equally post-hoc to inhibition or perturbation of the saccade-related network, or to strengthening of the fixation-related network. While “exploratory behavioural demonstrations" (Polanía et al., 2018) are required to establish that NIBS can induce behavioural effects, the high variability of behavioural effects following NIBS to the oculomotor system (Vernet et al., 2014) reinforces the need for neurophysiological investigation of the underlying mechanism.

### 4.3. State dependency and the dynamics of brain stimulation: insights from perturbing the oculomotor system

Saccadic measures provide a simple and accessible, albeit indirect means to assess the effects of NIBS on oculomotor targets, but they are only one component of the orienting response. As mentioned above, converging evidence demonstrates that the threshold for provoking an orienting response at the head is lower than that needed to provoke a saccade, and early work in the oculomotor system reported time-locked recruitment of neck muscles following the application of TMS to the human frontal cortex, including the FEF (Thickbroom et al., 1996). Subsequent work in both humans (Goonetilleke et al., 2015) and NHPs (Gu & Corneil, 2014) showed that such neck muscle responses exhibit state-dependency, being greater when the FEF is actively engaged at the time of TMS. Neck muscle recordings, and specifically the recruitment of a contralateral head turning synergy, may therefore provide access to a poly-synaptic, feedforward, and state-dependent oculomotor MEP that could parallel the use of the MEPs evoked by TMS-M1 (Baker et al., 1994; Barker et al., 1985) and serve as a positive control for the perturbation of brain activity via NIBS.

The issue of state dependency is critical for contemporary use of NIBS in humans: the contribution of a given brain area to a given task is often inferred from the ability of NIBS to influence behavioural output (Silvanto & Pascual-Leone, 2008). Recent work of ours using intracortical microstimulation showed that the availability of multiple measures of oculomotor output can illustrate the complexity of the neural responses to stimulation of a high-level area. We showed that sub-saccadic threshold stimulation of the supplementary eye fields (SEF), another important frontal oculomotor areas, can also evoke a neck muscle response in the absence of saccades (Chapman et al., 2012), consistent with a general role for this area in orienting. Furthermore, we found that larger neck muscle responses were evoked when sub-saccadic ICMS was delivered as NHPs prepared to make an anti- versus a pro-saccade (Chapman & Corneil, 2014), which is in line with previous work showing that the SEF is more active during antisaccades (Schlag-Rey et al., 1997). This is consistent with the idea of neck muscle responses providing a feedforward and state-dependent measure of SEF excitability at the time of stimulation onset; presumably, stimulation is summing with higher levels of endogenous SEF activity to create a larger neck muscle response. Importantly, and in marked contrast, SEF stimulation on these same anti-saccade trials delayed saccadic reaction times. In other words, two state-dependent output measures within the same experiment were oppositely affected by stimulation within the same trials (*Figure 2*).

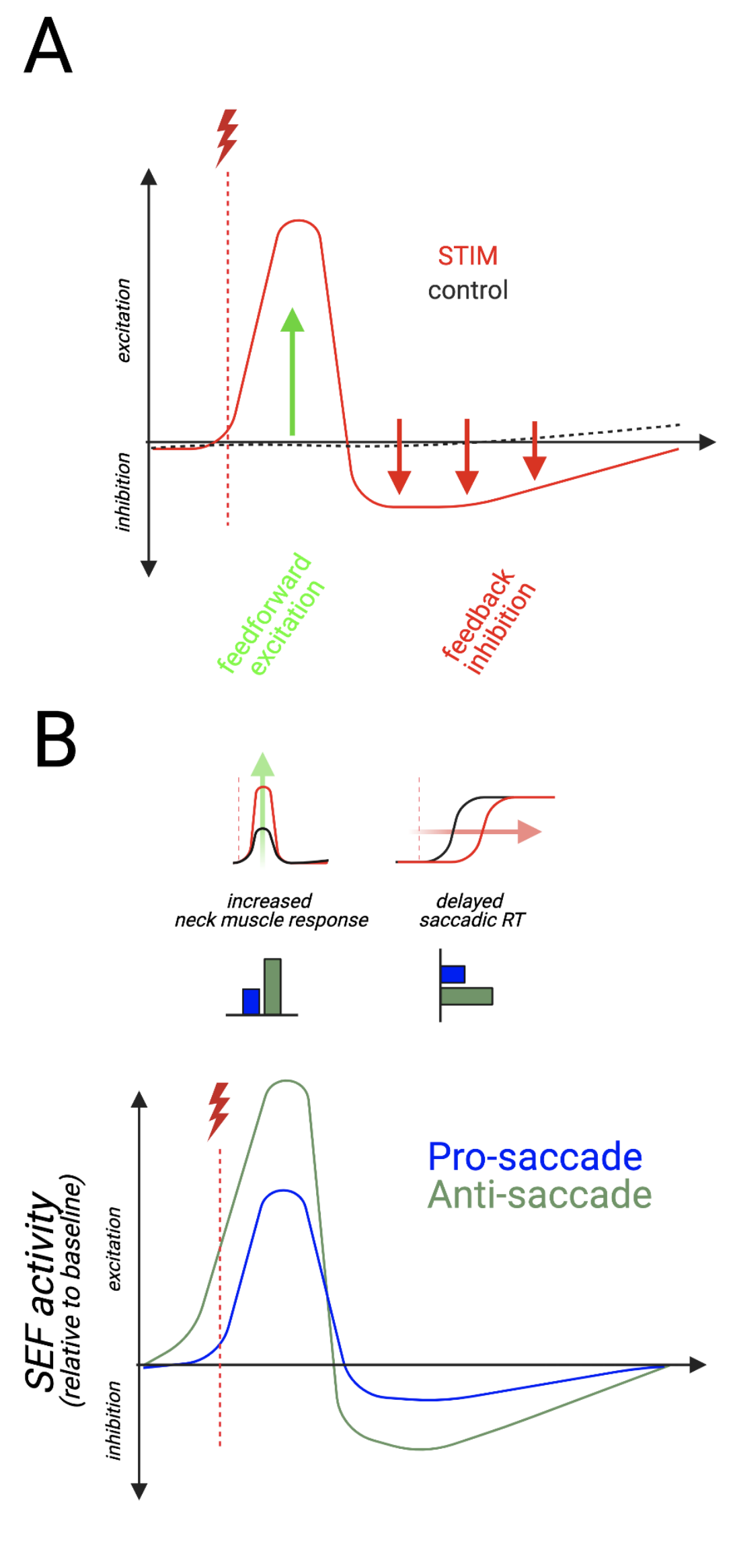

***Figure 2 - Excitation-Inhibition responses following brain stimulation. A.*** *Excitation-then-inhibition dynamics are commonly observed following invasive and non-invasive brain stimulation, in which a short phase of state-dependent feed-forward excitation is followed by a prolonged period of feedback inhibition.* ***B.*** *In the oculomotor system, such dynamics can produce a diversity of responses within a single trial. For example, ICMS of the supplementary eye fields induces a feed-forward excitation of neck-muscles on precisely those trials where stimulation prolongs saccadic reaction time. Such effects are state-dependent, as they are modulated by the endogenous neural activity within the SEF at the time of stimulation (see Chapman et al 2010)*

While perhaps counterintuitive at first, those results reveal some important principles about behavioural results following brain stimulation: it illustrates the shortcomings of straightforward interpretations of the brain's response to stimulation from a single behavioural measure. Had one only measured reaction times, one could have reasonably surmised that stimulation 'disrupted' the saccade network, 'enhanced' the fixation network, or somehow impaired oculomotor processing as a whole. Such interpretations are not consistent with a feed-forward, polysynaptic recruitment of an evoked neck muscle response. Instead, we interpret these results as attesting to the complexity of how an interconnected network responds to stimulation: differential output measures from the same trials can go in opposite directions because of the network's dynamic response to stimulation. In the case of these results, we surmise that the state-dependent feedforward excitation that elicits the neck muscle response is followed by a more prolonged period of feedback inhibition of the SEF that delays saccade generation. Such an excitation-then-inhibition dynamic is consistent with observations from the FEF that combined intracortical microstimulation and optical imaging (Seidemann et al., 2002). Whether the prolonged period of inhibition results from intrinsic circuits (e.g., within the FEF) or extrinsic circuits (e.g., from the SC back to the FEF via the medio-dorsal nucleus of the thalamus) remains to be determined.

Interestingly, similar excitation-then-inhibition dynamics have been reported in the few studies investigating the immediate effects of TMS on spiking activity of motor and parietal cortex of the behaving monkey (Mueller et al., 2014; Romero et al., 2019). While the response patterns tended to be reliable for individual neurons, both studies observed a diversity of responses between neurons, including both excitatory responses as well as more complex excitation-inhibition-excitation patterns following single TMS pulses (Romero et al., 2019). A brief period of excitation followed by prolonged inhibition was also reported in the macaque visual system following ICMS of the lateral geniculate nucleus (LGN), hinting at a disruption of cortico-cortical signal propagation by silencing the outputs of cortical areas whose affarents are stimulated (Logothetis et al., 2010), possibly largely mediated via cortico-subcortico-cortical pathways. This is further supported by the central role of the thalamus as a hub for relaying sensori-motor information, orchestrating the interactions between distributed cortical nodes and providing the substrate for cognitive processing (Shine, 2020). Together, these results hint at a critical role of subcortical nodes like the superior colliculus and thalamus in shaping the brain's dynamic response to stimulation, including visual and sensori-motor processing within the oculomotor network (Seung, 1996; Watanabe et al., 2014).

### 4.4. Towards a comprehensive understanding of brain stimulation: comparing invasive and non-invasive techniques

NHP studies also provide the opportunity to directly compare the effects following NIBS with an enormous body of research employing a variety of invasive techniques in sensori-motor networks and their effect on behaviour: besides stimulation via ICMS (Dominguez-Vargas et al., 2017; Ebitz & Moore, 2017; Ekstrom et al., 2009; Kagan et al., 2021; Logothetis et al., 2010; Moore & Fallah, 2004; Mushiake et al., 1999; Premereur et al., 2012), this includes temporary lesions induced by cryogenic (Chen et al., 2020; Ma et al., 2019; Peel et al., 2017; Takei et al., 2021) or pharmacological manipulation (Balan et al., 2019; Bogadhi et al., 2021; Dias & Segraves, 1999; Sommer & Tehovnik, 1997; Wardak, 2006; Wilke et al., 2012), permanent lesions (Adam et al., 2019; Schiller et al., 1987; Schiller & Chou, 1998), or optogenetics (Diester et al., 2011; Tremblay et al., 2019; H. Watanabe et al., 2020). Such results from experiments combining selective manipulation and neural recordings in well-studied networks highlight the complexity of causally linking perturbed activity to behaviour. As an example, we recently performed a series of experiments where we examined the effect of cryogenic inactivation of the FEF on oculomotor behaviour and SC activity (Peel et al., 2017). While FEF inactivation increased contralateral SRTs as expected, the key question was how changes in SC activity related to such SRT increases. Prominent models of saccade initiation in areas like the SC or FEF surmised that SRTs were related to the accumulation of neural activity toward a threshold (Everling et al., 1999; Heitz & Schall, 2012; Jantz et al., 2013; Paré & Hanes, 2003), predicting that SRT increases could be caused by one or both of an increase in threshold or a decrease in the rate of the accumulation of neural activity. Importantly, SC activity during FEF inactivation followed neither prediction; instead, SRT increases were best explained by delays in the time at which neural activity started to accumulate (Peel et al., 2017). This result echoes other work on how our conceptualizations of neural activity ultimately have to be validated, and perhaps modified, with recordings of neural activity to better understand the link between brain and behaviour (Heitz & Schall, 2012, 2013; Schall, 2019).

## 5. The role for NHP studies in understanding the brain's response to stimulation - future perspectives

The examples given above reinforce a core problem for the interpretation of brain stimulation results: the absence of a thorough understanding of the causal chain of events, starting with the activity directly induced by NIBS, through to within- and between-area effects which are themselves influenced by endogenous activity, to the production of behavioural output. Any form of stimulation forces an artificial and unnatural profile of activity whose influence can spread to interconnected nodes. While progress is being made in simulating and modeling cell-type specific and anatomical responses within the targeted area (Aberra et al., 2020), it is essential to extend this approach to other areas, and ultimately link such induced activity to behaviour. It is in bridging the network effects of NIBS to behaviour that we feel the NHP has the most critical role to play. Invasive recordings in NHPs can provide information at a much higher spatial and temporal resolution than in humans. The homologies in NHPs and humans of brain networks like the visual and oculomotor networks make it likely that the dynamic response to NIBS will be similar. Ongoing NHP work with other experimental techniques potentiates comparison of behavioural and neural results to those induced by NIBS, which in turn can lead to physiologically-informed tests of hypotheses and heuristics of the effects of NIBS. Recent progress in modelling and artificial neural networks to reconstruct and predict activity within visuo-motor networks have been used to mimic, simulate, and compare the neural and behavioural effects of perturbation (Kar et al., 2019; Michaels et al., 2020; Rajalingham et al., 2018; Yang et al., 2021), which might also prove useful in the development of closed-loop brain stimulation approaches. Emerging NIBS techniques like fUS (Kubanek et al., 2020; Verhagen et al., 2019; Yang et al., 2018, 2021) and TI (Grossman et al., 2017) can target subcortical structures like the thalamus, hippocampus, cerebellum, or basal ganglia; doing this in NHPs is essential given the growing appreciation of the role of subcortical structures in higher-order behaviour (Basso et al., 2021; Bogadhi et al., 2021). Finally, neuromodulatory therapies such as vagal nerve stimulation are being explored as a treatment mode for a variety of psychiatric and cognitive disorders, despite a lack of a mechanistic understanding of why this intervention can be effective. Again, the NHP offers a model system in which to build just such a mechanistic understanding of how such an intervention can impact higher-level behavior.

The proximity of NHPs to humans, which can be conceptualized in terms of phylogeny, brain circuitry, behavioural repertoire, or cognitive capabilities, positions the NHP model as a means to tackle fundamental questions in NIBS research. As with any animal model, caution is always warranted, and there are limitations. NHPs are not simply scaled-down humans, and they occupy a unique evolutionary niche for which their brain networks have evolved. Their motivating goal in a lab is different from humans, as NHPs perform in order to maximize reward. NHPs also learn tasks via a trial-and-error process that increments task complexity. For a variety of pragmatic reasons, and unlike the typical human NIBS study, NHP data is usually collected from highly-overtrained subjects. Such overtraining may reduce response variability compared to what is observed in humans. As NHPs cannot verbally report evoked responses such as phosphenes or other complex percepts, alternative approaches using two-forced choice alternatives (Murphey et al., 2009; Ni & Maunsell, 2010; Tehovnik et al., 2004) or more natural, unconstrained, and exploratory tasks are required to infer induced activity (Krause et al., 2017).

Another challenge fundamental to NHP work is that studies typically use between 2 and 4 animals. Such sample sizes are far lower than those in rodent or human studies, and are a potential concern given the heterogeneity of the effects of NIBS seen in humans. While the lower sample size is offset by repeating experimental sessions, doing so raises questions about optimal intervals between repeats to avoid carry-over effects. At the current time, there is no universally accepted schedule for how NIBS sessions should be planned, nor integrated with various control sessions, hence the onus is on the investigator to rationalize their choices, which themselves can be influenced by the idiosyncratic preferences of an animal. Given the inherent costs of NHP experiments, it is essential that null findings from NIBS-NHP experiments are reported in the literature in order to optimize collective efforts. Pre-registered studies, in which experimental schedules and controls are clearly defined, will also have a role. NIBS experiments in NHPs can be extremely time-consuming, and an open culture of transparency is needed to optimize collective efforts; doing so mirrors trends in the human literature (de Graaf & Sack, 2011; de Graaf & Sack, 2018).

A different issue relates to the suitability of NHP-based studies of NIBS to understand or treat higher-level cognitive disorders or conditions. While some disorders or conditions can be acquired or induced, such as stroke, others are either neurodegenerative conditions that manifest over many years (e.g., Alzheimer's or PD), or are arguably human conditions (e.g., depression) that may lack a direct macaque-based model. While there has been considerable progress in developing transgenic macaque models (for a review see Park & Silva, 2019), macaques have a comparatively slower breeding cycle that requires significant investment of both time and money to establish. Transgenic approaches can be more easily established in another emerging NHP model, the common marmoset (Callithrix jacchus), which itself can fill a gap between rodent and macaque models. The marmoset oculomotor system and their natural orienting behaviour resembles that of macaques (Schaeffer et al., 2020), hence this system is amenable to perturbation using either invasive or non-invasive techniques. Further, marmosets share key facets of natural social cognition and communication with humans, which may make them a better animal model for complex human social behaviours and the dysfunction of such behaviours in neuropsychiatric disorders (Miller et al., 2016).

The use of animals in research is governed by a variety of regulations that vary country-by-country, and approval for use requires adhering to numerous animal-specific ethical and welfare considerations. As with many countries, the Canadian Council on Animal Care emphasizes the “3Rs” (Replacement, Reduction, Refinement) in research. Our review here has emphasized the vital role that NHPs can play in advancing an understanding of NIBS, as this animal model is uniquely suited for linking the effects of brain stimulation to complex behavior. Thus, as the NHP model can offer answers that cannot be gained from other animal models, in our view they cannot be completely replaced. That being said, other animal models may well be more suited to answer other questions about NIBS, such as fundamental questions about biophysics and aspects of safety, and it is incumbent on the researcher to carefully consider which animal species, if at all, are required to answer a specific question.

From a refinement perspective, continued assessment and improvement of husbandry and experimental techniques is essential, and the transparent dissemination of such refinements, along with the promotion of the benefits accrued from research in NHPs, can alleviate public concerns associated with animal experiments (Mitchell et al., 2021). Along these lines, a recent longitudinal study tracking long-term markers of stress and inflammation did not find evidence for negative effects of many procedures and techniques used for sensorimotor neuroscience in awake, behaving macaques (Wegener et al., 2021). Continued refinement in how NIBS is delivered or behaviour is measured, for example using touchscreens, markerless tracking, or wireless stimulation and recording setups, may advance the field by leading to assessments of more natural behaviours, lessening the need for extensive training. It is incumbent on the experimenter to establish and use routines that are the least stressful to the animal; as emphasized in many reviews of macaque sensorimotor neuroscience, good science is predicated on a healthy animal (Prescott et al., 2010).

Finally, from a reduction standpoint, several initiatives from NHP facilities are strengthening international networks (e.g., EUPRIM-Net & PRIME-RE, PRIMatE-Resource Exchange) and creating open-source databases of translationally relevant brain data that should be expanded to those that incorporate NIBS (PRIMateE Data Exchange, or PRIME-DE, Milham et al., 2018; Vanduffel, 2018). It is our opinion that an emphasis on sharing and publishing negative results is particularly important for studies using NIBS in NHPs. Doing so can move the field toward best practices and avoid positive confirmation biases (de Graaf and Sack 2010), both of which can ultimately reduce the overall number of research subjects. Recent meta-studies have illustrated the benefits of such open sharing of information for improving the application of optogenetics and chemogenetics in NHPs (Tremblay et al., 2020).

The field of non-invasive brain stimulation is at a crossroads: despite a steadily increasing use in both basic research as well as therapeutic use in humans, we still lack a thorough understanding of the effects following NIBS techniques across multiple levels of neural architecture, as well as their subsequent impact on behaviour. Further establishing the non-human primate as a key model for NIBS offers the opportunity to investigate and better understand brain processes using non-invasive stimulation techniques. In doing so, this will further the informed application of NIBS techniques for the treatment of complex cognitive disorders.

**Abbreviations:**

cTBS continuous theta burst stimulation,

DBS deep brain stimulation

FEF frontal eye field(s)

fUS focused ultrasound

ICMS intracortical microstimulation

iTBS intermittent theta burst stimulation

LGN lateral geniculate nucleus

LIP lateral intraparietal cortex

M1 motor cortex

MEP motor evoked potential(s)

NHP non-human primate

NIBS non-invasive brain stimulation

TMS transcranial stimulation

OPN omnipause neurons

PD Parkinson’s disease

RT reaction time(s)

rTMS repetitive transcranial magnetic stimulation

SC superior colliculus

SEF supplemental eye field(s)

tACS transcranial alternating current stimulation

tDCS transcranial direct current stimulation

tES transcranial electric stimulation